\documentclass[letter]{aa} 

\usepackage{hyperref}
\usepackage{graphicx}
\usepackage{txfonts}
\begin{document}

   \title{Very late-time spectroscopy of SN 2009ip: Constraints on the ongoing H$\alpha$ emission}


   \author{Thallis Pessi
          \inst{1}
          \and
          Jose L. Prieto
          \inst{1,2}
          \and
          Luc Dessart
          \inst{3}
          }

    \institute{Instituto de Estudios Astrof\'isicos, Facultad de Ingenier\'ia y Ciencias, Universidad Diego Portales, Av. Ej\'ercito Libertador 441, Santiago, Chile\\ 
              \email{thallis.pessi@mail.udp.cl}
         \and 
            Millennium Institute of Astrophysics MAS, Nuncio Monsenor Sotero Sanz 100, Off. 104, Providencia, Santiago, Chile
        \and Institut d’Astrophysique de Paris, CNRS-Sorbonne Université, 98 bis boulevard Arago, F-75014 Paris, France}

   \date{Received 29 June 2023; accepted 7 August 2023}

 
  \abstract
   {The peculiar supernova (SN) 2009ip is an ambiguous event that spurred many questions regarding its true origins. Here, we present very late-time spectroscopic and photometric observations of SN 2009ip, obtained 9 years (3274 days) after the 2012B outburst.}
   {We analyze the H$\alpha$ emission still present in the very late-time spectrum of SN 2009ip. We also obtain photometric measurements in the $r$, $g,$ and $i$ bands.}
   {We obtained observations of SN 2009ip on 2021 September 10 with the IMACS instrument at the 6.5 m Magellan Baade Telescope, located at the Las Campanas Observatory.}
   {SN 2009ip was detected in the $r$, $g,$ and $i$ bands, with an absolute magnitude in $r$ band of $\sim -8.66$~mag. 
   We show that the source faded significantly since the last observations in these bands.
   We further show that the very late-time spectrum contains a persistent H$\alpha$ emission, although no other emission lines were detected. We measured a full width at half maximum (FWHM) of $930 \pm 40 \ \textrm{km s}^{-1}$ and luminosity of $\sim 8.0 \times 10^{37} \ \textrm{erg s}^{-1}$ for the H$\alpha$ emission. The luminosity decreased relatively slowly in comparison to the last observations and its fading rate is very similar to other long-living interacting transients, such as SN 2005ip.
   Finally, we conclude that although these properties could be consistent with a non-regular core-collapse SN, they may also be explained through non-terminal explosion scenarios.}
   {}


   \maketitle
%

\section{Introduction} \label{sec:intro}

Massive stars are known for producing very intense mass loss due to instabilities and energetic outbursts. If accumulated material around the star is dense enough, the stellar activity might generate observed events known as interacting transients, characterized by emission lines that are result from the interaction of the ejecta with the circumstellar material (CSM). If the transient is a terminal event (i.e., a true core-collapse), it can be classified as a Type IIn (H-rich CSM), Ibn (He-rich CSM), or Icn (C- and O-rich) supernova (SN). 
For some of these events, the dense CSM hides the underlying mechanism, making it difficult to decipher their true nature, namely: whether it is a SN or a violent eruption in a massive star. 
The interacting transient known as SN 2009ip is a famous example of this kind. Discovered in 2009, the event was classified as a Type IIn SN when it reached a magnitude of M$_R \sim -14.5$ mag \citep{2009CBET.1928....1M}. Subsequent eruptions observed in the next couple of years proved that the source was still active and exhibited a behavior similar to erupting luminous blue variables \citep[LBVs,][]{2010AJ....139.1451S, 2011ApJ...732...32F, 2013ApJ...767....1P}. Very early images obtained by the Hubble Space Telescope (HST) in 1999 showed a very luminous progenitor at the position of SN 2009ip, with an initial mass of $50 - 80 \ \textrm{M}_\odot$, consistent with a LBV star \citep[][]{2010AJ....139.1451S}.

In 2012, two extreme events were observed in SN 2009ip: the first, named 2012A, reached a magnitude of  M$_V \sim -14.5$ mag, and the second, 2012B, reached M$_R \sim -18$ mag \citep{2013MNRAS.430.1801M, 2013ApJ...767....1P, 10.1093/mnras/stt813}. During the 2012B event, the H$\alpha$ profile equivalent width (EW) varied between $100$~\AA, at the beginning, to $1000$~\AA, $\sim 260$ days later, and presented a variety of velocity components \citep[][]{2014MNRAS.442.1166M}. The full width at half maximum (FWHM) of the H$\alpha$ narrow component went from $\sim 700 - 1000 \ \textrm{km s}^{-1}$ over this period, while the broad component showed a range of $\sim 10600 - 15000 \ \textrm{km s}^{-1}$. The blue absorption component of the H$\alpha$ P-Cygni profile also showed high velocities, extending up to $\sim 15000 \ \textrm{km s}^{-1}$ \citep[][]{2014MNRAS.442.1166M}.
 
These eruptions were followed by extensive discussion, and their true nature is still under debate \citep{2013ApJ...763L..27P,2014AJ....147...23L, 2014ApJ...780...21M, 2014MNRAS.438.1191S, 2014ApJ...787..163G, 2014MNRAS.442.1166M, 2015ApJ...803L..26M, 2013MNRAS.433.1312F, 2015MNRAS.453.3886F, 2015ATel.8417....1T, 2016MNRAS.463.2904S, 2017MNRAS.469.1559G}. 
Some authors claimed that the observed high velocities during 2012B were only possible if SN 2009ip was in fact a true SN \citep[e.g.,][]{2013MNRAS.430.1801M, 2014MNRAS.438.1191S}. 
On the other hand, \cite{2013ApJ...767....1P} proposed that such high velocities were also detected during the LBV-like variability observed in the years before, and \cite{10.1093/mnras/stt813} argued that the lack of nebular emission from nucleosynthesis material after the 2012B eruption could support the non-terminal scenario. 
\citet[][]{2015ApJ...803L..26M} demonstrated that the light curve of SN 2009ip after 2012B could be fitted by a CSM interaction model with $\sim 2 \times 10^{49} \ \textrm{erg}$, thus being not a regular SN explosion \citep[see also][for a discussion on spectroscopic modeling of Type IIn SNe]{2016MNRAS.458.2094D}.
Furthermore, the case for LBVs being the progenitors of SNe is still contested, namely:\ the case for SN 2020tlf and its pre-SN activity with a luminosity of $10^6$~L$_\odot$ and consistent with a $60$~M$_\odot$ progenitor; however, based on the light curve and spectra, the progenitor is consistent with a $12$~M$_\odot$ star \citep[][]{2022ApJ...924...15J}.

Late-time observations of SN 2009ip showed that the source had dimmed significantly after two years, reaching $g \sim 21.3$ on 2014 July 28 and $r \sim 19.7$ on 2014 December 16 \citep{2017MNRAS.469.1559G}. Spectroscopy obtained by \citet{2017MNRAS.469.1559G} on 2015 July 16 shows that SN 2009ip was still dominated by hydrogen in emission 1026 days after event B, with the presence of He I, Fe I, and the emission forest of Fe II. 
More recently, \citet[][]{2022MNRAS.515...71S} presented new photometry of SN 2009ip, obtained with the HST in December of 2021. They show that the source faded significantly, but steadily since 2013 and that now it is below the 1999 progenitor magnitude in the F606W band.
Here, we present late-time spectroscopic and photometric observations of SN 2009ip obtained on September of 2021, namely, 3274 days after the beginning of event B.

\section{Observations} \label{sec:obs}

We obtained optical spectroscopy and photometry of SN 2009ip on 2021 September 10 UT\footnote{UT dates are used throughout this paper}, using the IMACS instrument (+ f/2 camera) at the 6.5 m Magellan Baade Telescope. The spectra were obtained using long-slit observations with a 300~l/mm grism, the FourS\_2471 slit mask, and a wavelength coverage of $4200 - 9400$~\AA. The final spectrum was the result of the median combining of three exposures, with a time of integration of $1800$~s each.

The combined spectrum was reduced using the {\tt\string kosmos}\footnote{ \url{https://github.com/jradavenport/kosmos}} pipeline, following standard procedures of long-slit spectroscopic reduction, including bias and flat-field correction, wavelength, and flux calibration. We carried out an absolute flux calibration on the 1D spectrum using the $r$-band photometry and synthetic photometry obtained with the {\tt\string pyphot}\footnote{ \url{https://mfouesneau.github.io/pyphot/}} package. 

Imaging in the SDSS $gri$ filters was also obtained with IMACS f/2 camera. A total exposure time of $3 \times 90$~s was used in $r-$band, $1 \times 300$~s in $g-$band, and $2 \times 200$~s in $i-$band. The individual exposures in each band were reduced using standard procedures of bias subtraction and flat-field correction in the {\tt\string kosmos} pipeline. Astrometry correction was performed through {\tt\string Astrometry.net} \citep[][]{2010AJ....139.1782L} and the images were combined using {\tt\string SWarp} \citep[][]{2010ascl.soft10068B}. The photometry of SN2009ip was extracted using the point spread function fitting (PSF) package {\tt\string DAOPHOT} \citep[][]{1987PASP...99..191S, 2011ascl.soft04011S}. The photometric calibration was obtained using stars in the field with Pan-STARRS1 \citep{2002SPIE.4836..154K, 2016arXiv161205560C, 2020ApJS..251....7F} photometry. The result of our photometry in the $gri-$bands is reported in Appendix \ref{app:phot}.

SN 2009ip is located in the outskirts of the spiral galaxy NGC 7259, at $\alpha = 22^h23^m08.261^s$ and $\delta = -28^{\circ}56'52.40"$ (J2000.0). In this work, we use a distance modulus of $\mu \approx 32.0$, a redshift of $z = 0.00594$ \citep[][]{2014ApJ...787..163G, 2017MNRAS.469.1559G} and a Galactic extinction of $A_V \approx 0.054$ \citep[][]{2011ApJ...737..103S}.

\section{Results and discussion} \label{sec:res}

\begin{figure*}[t!]
\centerline{\includegraphics[scale=0.75]{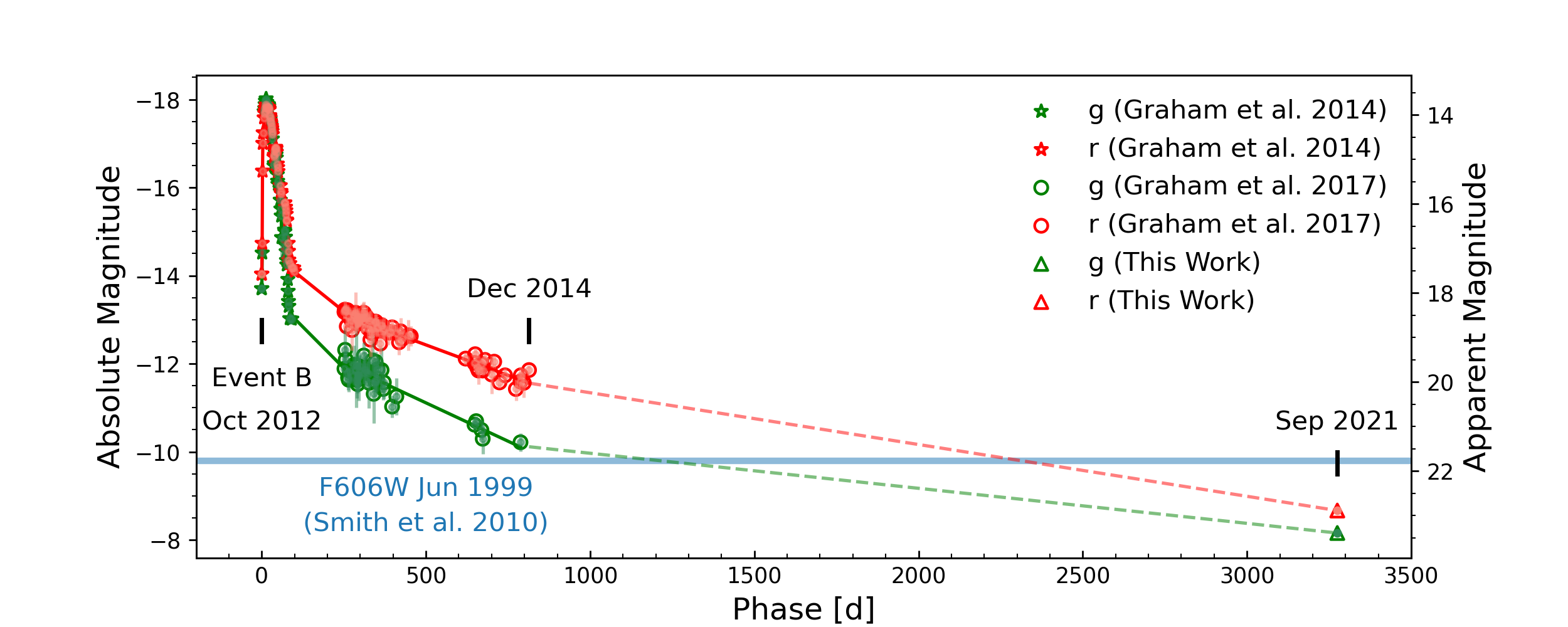}}
\caption{Late-time photometric evolution of SN 2009ip, with the photometry presented in this work in $r$ (red triangle) and $g$ (green triangle) bands, obtained on 2021 August 10. The figure also shows the photometry in $r$ and $g$ bands around the event 2012B, presented in \citet[][]{2014ApJ...787..163G}, and the late-time photometric evolution in these bands, between 251 and 813 days after event B, presented in \citet{2017MNRAS.469.1559G}. the dashed lines show the inferred slope of the source between the last reported observation and our photometry. Phase is given in relation to the begging of event B. \label{fig:photometry_comparison}} 

\end{figure*}

\begin{figure*}[t!]

\includegraphics[scale=0.6]{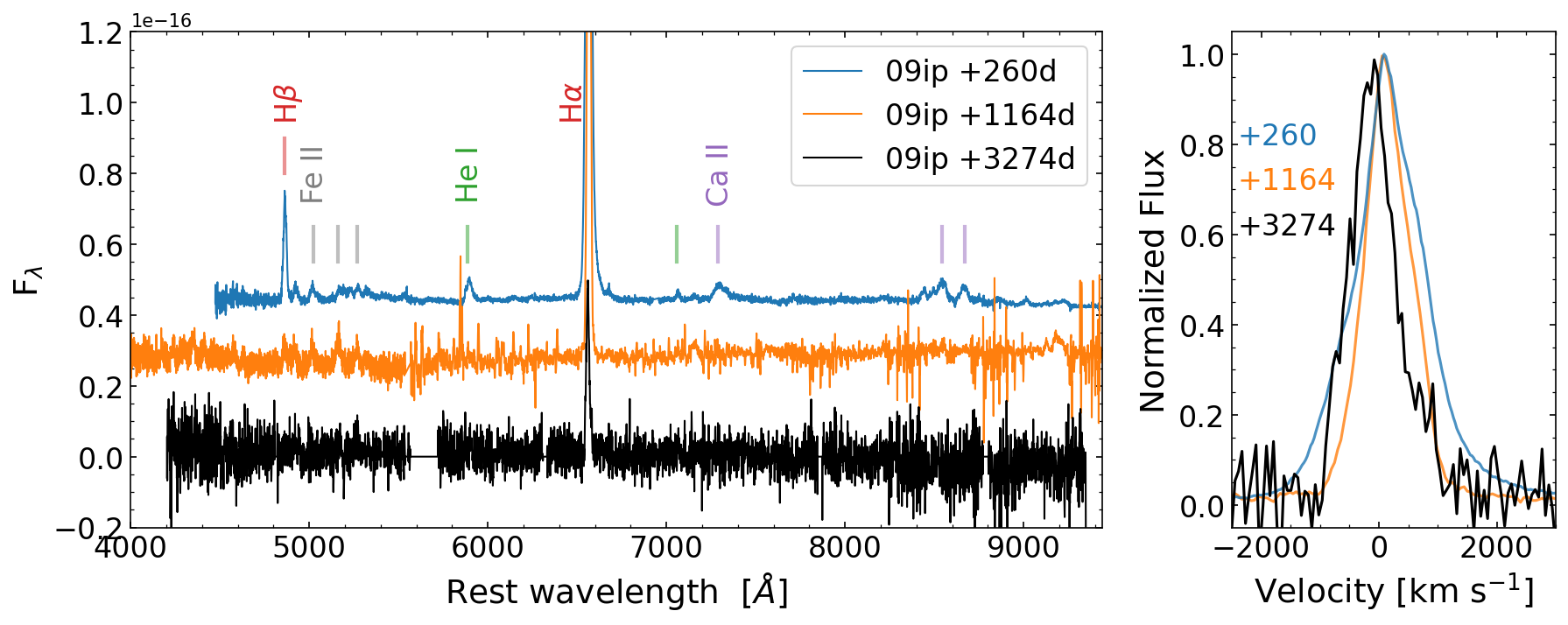}
\caption{ Very late-time spectrum of SN 2009ip obtained on 2021 September 10, or at phase +3274 relative to the event 2012B, obtained with the IMACS instrument at the 6.5 m Magellan Baade Telescope (in black). 
The spectrum is available on WISeREP\protect\footnotemark \citep[][]{2012PASP..124..668Y}.
We also show the last published spectrum of SN 2009ip from \citet{2017MNRAS.469.1559G}, obtained in 2015 December 11 (at phase +1164 days) and a spectrum from 2013 June 10 (at phase +260). \label{fig:spec_comparison}}

\end{figure*}

\begin{figure}[h!]
\centerline{\includegraphics[scale=0.7]{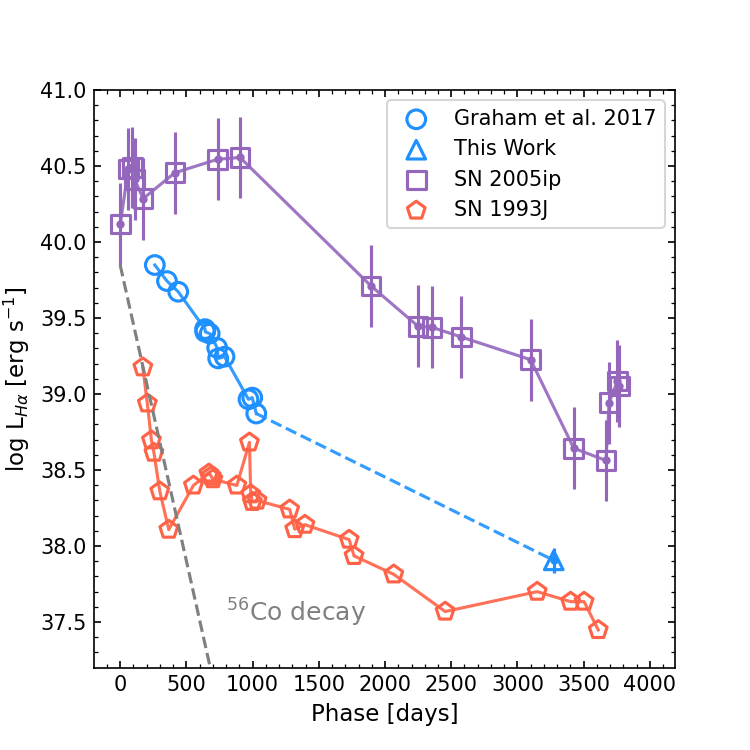}}
\caption{Luminosity evolution of the H$\alpha$ line in SN 2009ip between 251 and 3274 days after the peak of event B (in blue). The triangle shows the value presented in this work, while the circles represent the values from \citet{2017MNRAS.469.1559G}. We also show the evolution of the H$\alpha$ luminosity for the long-lived SN 2005ip \citep[purple squares, ][]{2009ApJ...695.1334S, 2017MNRAS.466.3021S}, between 1893 and 3768 days after peak brightness, and SN 1993J \citep[orange pentagons, ][]{2009ApJ...699..388C}, between 170 and 3608 days after peak. The gray dashed line highlights the expected decline rate of the luminosity if the H$\alpha$ emission was driven by $^{56}$Co decay, following $L \propto 2^{-t/77.3}$. \label{fig:L_evolution}} 

\end{figure}

\subsection{Photometry}

In Appendix \ref{app:fov}, we show the observed field of view in $r-$band, obtained on 2021 September 10, with an inset in a region close to the galaxy NGC 7259 showing a bright source at the position of SN 2009ip. We measured a magnitude of $r=22.93 \pm 0.08$~mag; or a Galactic extinction-corrected absolute magnitude of M$_r = -8.66 \pm 0.08$~mag. The source was also detected in $g$ and $i$ bands, with $g = 23.46 \pm 0.07$~mag (M$_g = -8.15 \pm 0.07$~mag) and $i = 23.59 \pm 0.08$~mag (M$_i = -8.12 \pm 0.08$~mag). 

SN 2009ip faded significantly since the last reported detection of the source in these bands by \citet{2017MNRAS.469.1559G}, obtained on 2014 December 16 (shown in Figure \ref{fig:photometry_comparison}). Our photometric measurements show a fading of $\approx 3.19$~mag in $r-$band and of $\approx 2.07$~mag in $g-$band, in relation to the last observations. The fading rate of $\sim 0.0011 \ \textrm{mag day}^{-1}$ in $r$ band and $\sim 0.0007 \ \textrm{mag day}^{-1}$ in $g$ band implies a slowing in comparison to the rate estimated by \citet{2017MNRAS.469.1559G} at the time ($\sim 0.0030 \ \textrm{mag day}^{-1}$ and $\sim 0.0033 \ \textrm{mag day}^{-1}$, respectively). 
The change in the slope can be clearly seen in Figure \ref{fig:photometry_comparison}. Recently, \citet[][]{2022MNRAS.515...71S} reported the photometry of SN 2009ip obtained with the HST, and showed that all bands have a slow decline rate at very late times, in agreement with our work. Such decline rate is much slower than expected for radioactively powered emission and for normal Type II SNe \citep[although $^{56}$Co decay might not be the main powering mechanism for Type II SNe at late times,][]{2023arXiv230109089D}. 
However, this is commonly observed for other interacting transients that presented late-time emission generated by ejecta-CSM interaction \citep[e.g., 2005ip, ][]{2009ApJ...695.1334S, 2017MNRAS.466.3021S, 2020MNRAS.498..517F}.

One important point of comparison for the photometry is the pre-explosion detections of the progenitor of SN 2009ip. In Figure \ref{fig:photometry_comparison}, we show the magnitude of $\sim 21.8$ mag measured by \citet[][]{2010AJ....139.1451S} using the HST F606W filter on 1999 June 29.
The source is below this point in both $g$ and $r$ bands, which indicates that the source faded below its progenitor luminosity. This is in agreement with the results presented in  \citet[][]{2022MNRAS.515...71S}, who showed that SN 2009ip had faded considerably below its progenitor luminosity of 1999. The low luminosity in $gri$ bands does not necessarily imply a dead progenitor, as the surviving star could be emitting most of its flux in the bluer bands. A high UV flux was detected by the observations made by \citet[][]{2022MNRAS.515...71S}, which also points to a strong possibility of contamination from an underlying stellar cluster in that band.  

\citet[][]{2010AJ....139.1451S} also reported the source varying between $R \sim 20.37$, on 2005 July 03, and $R >21.51$ on 2008 August 23. The reason for such variability before a giant eruption is still not known, but such behavior is common in LBV stars \citep[][]{2010AJ....139.1451S} and is consistently observed for other 2009ip-like events. We note that the source does not appear to have produced any similar variability for the last 10 years, which is also supported by the photometry from \citet[][]{2022MNRAS.515...71S}. 
This could favor the scenario of event B of SN 2009ip as a true core-collapse event, since the source significantly dimmed its activity after a long period of considerable variability. However, this behavior would also be consistent in the scenario of pulsational pair-instability \citep[PPI, ][]{2017ApJ...836..244W} or a Roche lobe overflow (RLOF) -- that is, if a bare core was left after all the H-rich envelope being expelled during event 2012A. The mass loss could have moved the progenitor away from the PPI or RLOF regime, ceasing the violent bursts and variability that had previously been observed.

\subsection{Spectroscopy}

The spectrum of SN 2009ip obtained on 2021 September 10 (at phase +3274 relative to the beginning of event 2012B) is shown in Figure \ref{fig:spec_comparison}, where we also show the last published spectrum from \citet{2017MNRAS.469.1559G}, obtained on 2015 December 11 (at phase +1164 relative to the event 2012B), and an earlier spectrum, obtained on 2013 June 10 (at phase +260).
Although many different spectral features are still identified in the previous late-time spectrum of SN 2009ip, especially the lines of He I, Ca II, and Fe II, the spectrum presented here only shows one distinguishable emission feature of H$\alpha$.
However, the detection of SN 2009ip in the $g$ and $i$ bands,  could be caused by the presence of a faint continuum or by the presence of emission lines that are undetected in our spectrum. \footnotetext{\url{https://www.wiserep.org/}}

In the right panel of Figure \ref{fig:spec_comparison}, we show a comparison of the H$\alpha$ emission of the spectrum obtained 3274 days after event B, to the line present in the spectra observed at phases $+1164$ days and $+260$ days. SN 2009ip presented an increase of the H$\alpha$ equivalent width (EW) during the first year after the beginning of event B, going from $\sim 100$~\AA \ at the peak of that event to $\sim 1000$~\AA \ in 2013. In 2015, the  H$\alpha$ EW had increased to about $\sim 2400$~\AA. 
Although we cannot estimate the H$\alpha$ EW in our spectrum, due to the continuum flux being close to zero, we note that the line width did not present a strong change from 2015. 
This EW is considerably larger than the observed for Eta Car \citep[$\sim 400 - 600$~\AA, ][]{2003ApJ...586..432S, 2005A&A...435..303S} and quiescent LBV stars, but commonly observed for Type IIn SNe and other interacting events \citep[][]{2014MNRAS.438.1191S}. 
This suggests a scenario of the H$\alpha$ emission being generated from the interaction of the remaining ejecta of event B with the CSM -- and not from the LBV-like ionizing mechanism of a survivor star. 

By using a Gaussian model, we fit the H$\alpha$ emission line and obtain a FWHM of $930 \pm 40 \ \textrm{km s}^{-1}$ and a luminosity of $8.0 \pm 0.7 \times 10^{37}  \ \textrm{erg s}^{-1}$. 
The FWHM is similar to the velocity measured in the spectrum obtained at phase $+1164$~d ($\sim 950$~km~s$^{-1}$), showing no evolution with time.
Figure \ref{fig:L_evolution} shows the evolution of the H$\alpha$ emission profile in SN 2009ip, with measurements presented by \citet{2017MNRAS.469.1559G} between phases $+251$~days and $+1027$~days. The luminosity of the line decreased steadily over the years and dimmed significantly in comparison to the luminosity presented in \citet{2017MNRAS.469.1559G}: the luminosity decreased by about two orders of magnitude between phases $+1027$~days and $+3274$~days, going from $\textrm{log} \ (L_{H\alpha}) \sim 40 \ \textrm{erg s}^{-1}$ to $\textrm{log} \ (L_{H\alpha}) \sim 38 \ \textrm{erg s}^{-1}$. 

In Figure \ref{fig:L_evolution}, we also show the evolution of the H$\alpha$ luminosity for two long-lived interacting transients, SN IIn 2005ip \citep[][]{2009ApJ...695.1334S, 2017MNRAS.466.3021S, 2020MNRAS.498..517F} and the SN IIb 1993J \citep[][]{2009ApJ...699..388C}, and a comparison with the expected decline rate of luminosity if the line was powered by $^{56}$Co decay, following a $L \propto 2^{-t/77.3}$ behavior \citep[however, we note that $^{56}$Co decay might not be the main powering source of normal Type II SNe at very late times,][]{2023arXiv230109089D}. These two SNe have been followed up for a very long period after peak, with their H$\alpha$ flux evolution being detected up to 3768 days and 3608 days after peak, respectively. Often the scenario of an extended mass loss in a red supergiant (RSG) is used to explain the long-lived emission of SN 1993J and other similar events \citep[see, e.g.,][]{2012MNRAS.424.2659M, 2012ApJ...751...25M, 2015MNRAS.449.1876S}, with an estimate of strong wind with $10^{-4} \ \textrm{M}_\odot \ \textrm{yr}^{-1}$  \citep[][]{2012MNRAS.424.2659M}. SN 2005ip remained at a much larger luminosity, requiring a more extreme progenitor mass loss rate to explain its increased late-time luminosity. 
\citet[][]{2017MNRAS.466.3021S} shows that such luminosities are consistent with a progenitor mass loss rate of $10^{-3} \ \textrm{M}_\odot \ \textrm{yr}^{-1}$, which agrees well with the detected emission in radio and X-rays and requires a long-lived pre-explosion mass loss. 
Recent analyses estimated the pre-explosion mass loss rate of SN 2005ip to be of $10^{-2} \ \textrm{M}_\odot \ \textrm{yr}^{-1}$ and to have produced a total mass loss of $\geq 1 \ \textrm{M}_\odot$ just prior to explosion \citep[][]{2020MNRAS.498..517F}.
We note that SN 2009ip shows a very similar fading behavior of the H$\alpha$ luminosity to these two SNe; therefore, it is consistent with a late-time interaction between the ejecta and an extended CSM (although we note that the mechanism behind SN 1993J is different than SN 2005ip and SN 2009ip and that it presents a very distinct velocity evolution, with a very broad H$\alpha$ profile and velocities up to $\sim 5000 - 10 000$~km~s$^{-1}$). 

The overall luminosity in SN 2009ip is about 1.5 orders of magnitude lower than SN 2005ip, suggesting that the amount of mass lost during the event B of SN 2009ip should be around ten times less than for SN 2005ip.
Such a smaller luminosity, however, could also be caused by the geometry of the event, as a disk-like geometry would require more energy to emit the same amount of luminosity as observed in a spherical CSM. 
As can be seen in Figure \ref{fig:spec_comparison}, the H$\alpha$ emission has the peak of the line shifted to $v \sim - 100 \ \textrm{km s}^{-1}$. 
Asymmetries in the H$\alpha$ emission were noticed at early-times \citep[][]{2014ApJ...780...21M, 2017MNRAS.469.1559G}, which suggested a non-spherical CSM around the explosion.
However, the shift in the spectrum presented here might not be significant. This and the fact that the previous late-time spectra show a symmetric profile of H$\alpha$ make the scenario of a non-spherical CSM less likely.

Finally, given the low velocities, it is also possible that the very-late spectrum reported here originated from the inner ejecta of SN 2009ip. If that is the case, this suggests that there are no metals in the emitting region, making the case for a core-collapse event harder to explain and more consistent with a low-metallicity non-terminal explosion. However, more detailed observations of the inner regions are needed to confirm this scenario.

\section{Summary and conclusions} \label{sec:conc}

In this work, we presented very late-time photometry and spectroscopy of SN 2009ip, obtained on 2021 September 10, namely, 3274 days after the event 2012B. We show that the source faded significantly since the last observations obtained in $g$ and $r$ bands in 2014. The slow fading rate is comparable to long-lived SNe and other interacting events. 
The source appears to have been fading steadily since 2014, with no strong variability or explosive outbursts like the ones observed in the years before the main events of 2012. 
However, this could also be consistent with a scenario of a PPI or RLOF event.

We show that the very late-time spectrum of SN 2009ip shows a persistent narrow H$\alpha$ emission feature, with a luminosity of $\sim 8 \times 10^{37}  \ \textrm{erg s}^{-1}$ and an FWHM of $\sim 900 \ \textrm{km s}^{-1}$. The H$\alpha$ luminosity evolution is very similar to other transients with late-time CSM interaction. The H$\alpha$ luminosity of SN 2009ip falls between the luminosities for SN 2005ip and SN 1993J, although it has a different velocity evolution than these two SNe.
The similarity of the H$\alpha$ EW to the value observed at the end of event B might indicate that the line is being produced by the still interaction of the ejecta with the CSM, being the result of the deceleration of the explosion ejecta or the ionization of the shocked CSM. 
Another possibility is that the spectrum is generated by the inner ejecta of SN 2009ip, a case that is supported by the low velocities reported here. If that is the case, the fact that no metals are detected makes the scenario of a core-collapse of SN 2009ip less likely.

After 9 years of event B, SN 2009ip is still a dubious and hard-to-explain event. We showed that it is still hard to confirm if it was a true core-collapse supernova event, as other possibilities might explain the 2012 explosions.
The only certainty is that the mechanisms behind SN 2009ip are not the same as a regular Type IIn SN.
Future late-time observations and modeling of SN 2009ip and similar events might help in understanding the mechanism behind this ambiguous event.

\begin{acknowledgements}

T.P. acknowledges the support by ANID through the Beca Doctorado Nacional 202221222222.
J.L.P. acknowledges support by ANID through the Fondecyt regular grant 1191038 and through the Millennium Science Initiative grant ICN12\_009, awarded to The Millennium Institute of Astrophysics, MAS.
This work makes use of the following softwares:
kosmos (\url{https://github.com/jradavenport/kosmos}), 
Astropy \citep{astropy:2013, astropy:2018}, 
SWarp \citep[][]{2010ascl.soft10068B},
Astrometry.net \citep[][]{2010AJ....139.1782L},
DAOPHOT \citep[][]{1987PASP...99..191S, 2011ascl.soft04011S},
SUPERBOL \citep[Version 1.7;][]{2018RNAAS...2..230N} and
Pyphot (\url{https://mfouesneau.github.io/pyphot/}).

\end{acknowledgements}

\bibliography{ref}
\bibliographystyle{aasjournal}

\begin{appendix} 

\section{Photometry}\label{app:phot}

Table \ref{tab:phot} reports the $gri$ photometry of SN 2009ip, obtained on 2021 September 10 UT.

\begin{table*}
\caption{IMACS photometry of SN 2009ip\label{tab:phot}}
\centering
\begin{tabular}{cccccccc}
\hline
UT Date & MJD & \multicolumn2c{g} & \multicolumn2c{r} & \multicolumn2c{i}\\
 & & (mag) & (error) & (mag) & (error) & (mag) & (error)\\
\hline
2021-09-10 & 59468 & 23.46 & 0.07 & 22.93 & 0.08 & 23.59 & 0.08\\
\hline
\end{tabular}
\end{table*}

\section{Observed field of view}\label{app:fov}

\begin{figure*}[t!]
\centerline{\includegraphics[scale=0.9]{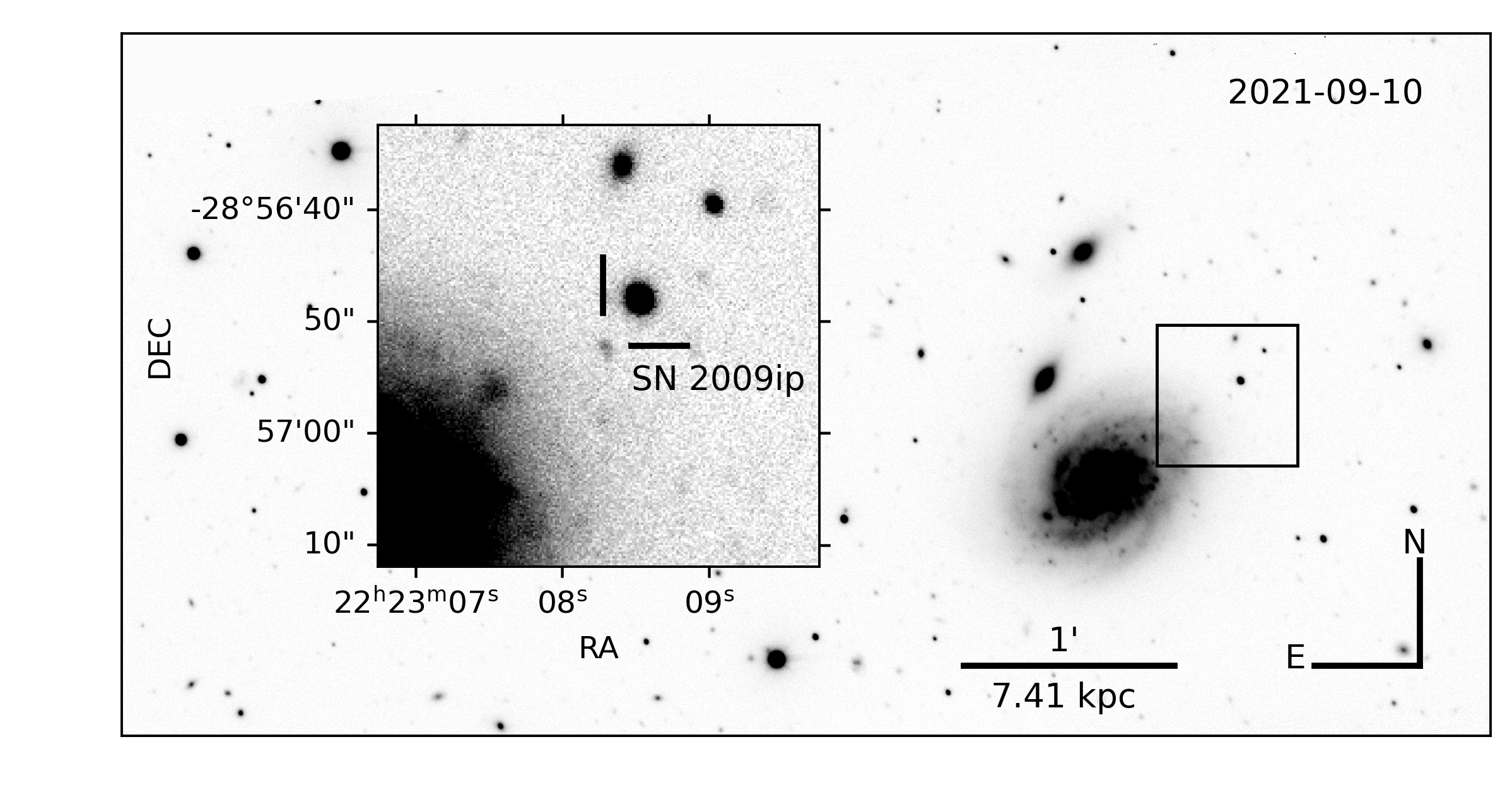}}
\caption{Observed field of view of the galaxy NGC 7259 in $r$ band, obtained with the IMACS instrument at the 6.5 m Magellan Baade Telescope, on 2021 September 10. The inset shows a bright source at the position of SN 2009ip. The $1\arcmin$ horizontal bar corresponds to $7.4$~kpc at a distance of $25.5$~Mpc. The resultant photometry is shown in Table \ref{tab:phot}. \label{fig:r_band_2009ip}} 

\end{figure*}

Figure \ref{fig:r_band_2009ip} shows the observed field of view in $r-$band of the galaxy NGC 7259, obtained on 2021 September 10, with an inset at the position of SN 2009ip.

\end{appendix}
\end{document}